\documentclass[12pt]{iopart}

\usepackage{iopams}
\usepackage[dvips]{graphicx}

\begin{document}

\title[The soundscape dynamics of human agglomeration]{The soundscape dynamics of human agglomeration}

\author{Haroldo V Ribeiro$^{1,2}$, Rodolfo T de Souza$^{1}$, Ervin K Lenzi$^{1,2}$, Renio S Mendes$^{1,2}$ and Luiz R Evangelista$^{1}$}
\address{$^1$ Departamento de F\'isica, Universidade Estadual de Maring\'a, Av. Colombo 5790, 87020-900, Maring\'a, PR, Brazil}
\address{$^2$ National Institute of Science and Technology for Complex Systems, CNPq, Rua Xavier Sigaud 150, 22290-180, Rio de Janeiro, RJ, Brazil}
\ead{hvr@dfi.uem.br}

\begin{abstract}
We report a statistical analysis about people agglomeration soundscape. Specifically, we investigate the
normalized sound amplitudes and intensities that emerge from people collective meetings. Our findings
support the existence of nontrivial dynamics characterized by heavy tail distributions in the sound 
amplitudes, long-range correlations in the sound intensity and non-exponential distributions in
the return interval distributions. Additionally, motivated by the time-dependent
behavior present in the volatility/variance series, we compare the observational data with those obtained from
a minimalist autoregressive stochastic model, a GARCH process, finding a good agreement.
\end{abstract}

%89.65.-s	Social and economic systems
%89.75.-k	Complex systems
%05.45.Tp	Time series analysis
%89.20.-a	Interdisciplinary applications of physics

\maketitle
\section{Introduction}
Physicists are now addressing problems very far from their traditional domain. Even social phenomena
are now ubiquitous in the research made by the statistical physicists~\cite{Castellano}. In particular,
the general framework of the Statistical Physics has been successfully applied to diverse interdisciplinary
fields ranging from finance~\cite{Vandewalle}, genetics~\cite{Peng} and biology~\cite{Berg}, to religion~\cite{Picoli}, 
tournaments~\cite{Ribeiro}, culinary~\cite{Kinouchi}, and music~\cite{Correa}. 

Naturally, in social phenomena the basic constituents of the system is the human. Humans are known
to have nontrivial collective dynamics, much more complicated than idealized physical interacting systems.
Moreover, even individual aspects related to social agents may not be available. This complex scenario is
reflected, in some sense, in several human activities. For instance,  elections~\cite{Fortunato}, collaborations between
actors~\cite{Watts} and also between scientists~\cite{Newman}, phone text-message~\cite{Zhou}, mail~\cite{Oliveira,Vazquez} or email~\cite{Vazquez,Barabasi} communication,
human travel~\cite{Brockmann,Gonzalez}, and collective listening habits~\cite{Lambiotte, Lambiotte2} are just a few examples where complex structures have been found.

Most of the previous investigation deal with record data obtained directly or indirectly from the system, trying to extract
some patterns or regularities about the system dynamics. This approach has been a trend towards investigating
social phenomena and also complex systems in general~\cite{Auyang,Jensen,Barabasi2,Boccara,Sornette}. Within this framework the most diversified
data were used as the sound. ``Listen to'' the system dynamics may be both a simple task and a minimally 
invasive measurement.  In this direction, several studies focused on the sound time series have been done.
Just to mention a few: researches about the acoustic emission from crumpled paper~\cite{Kramer,Mendes}, from paper fracture~\cite{Salminen}, 
and fractures in general~\cite{Sethna,Minozzi} show several features 
related to critical phenomena, the power spectrum of music and speech sounds presents $1/f$-like spectra~\cite{Voss} and
the normalized sound amplitude shows non-Gaussian features~\cite{Mendes2}, traffic flows were investigated by using the 
sound noise revealing scaling and memory~\cite{Skagerstam}, avalanches-like dynamics
was found in the sound of popping bubbles in foams~\cite{Vandewalle2} and also in the lung sound~\cite{Alencar}.

In this work, we present an investigation about a very common situation related to human collective activities: the people
agglomeration. Human beings agglomerations can emerge in various places for different reasons, for example, people having lunch 
in restaurant, parties, and working meetings. In all these situations a common and notorious feature is perceptible: the resulting
sound noise from these agglomerations. Here, our main goal is to show that a nontrivial dynamic emerges when analyzing
this kind of time series. In addition, employing a minimalist model we are able to reproduce statistical aspects of the empirical data. 
In the following, we present the details about the data acquisition, the statistical analysis of the data,
our minimal model, and finally we end with a summary. 
\begin{figure}[!ht]
\centering
\includegraphics[scale=0.46]{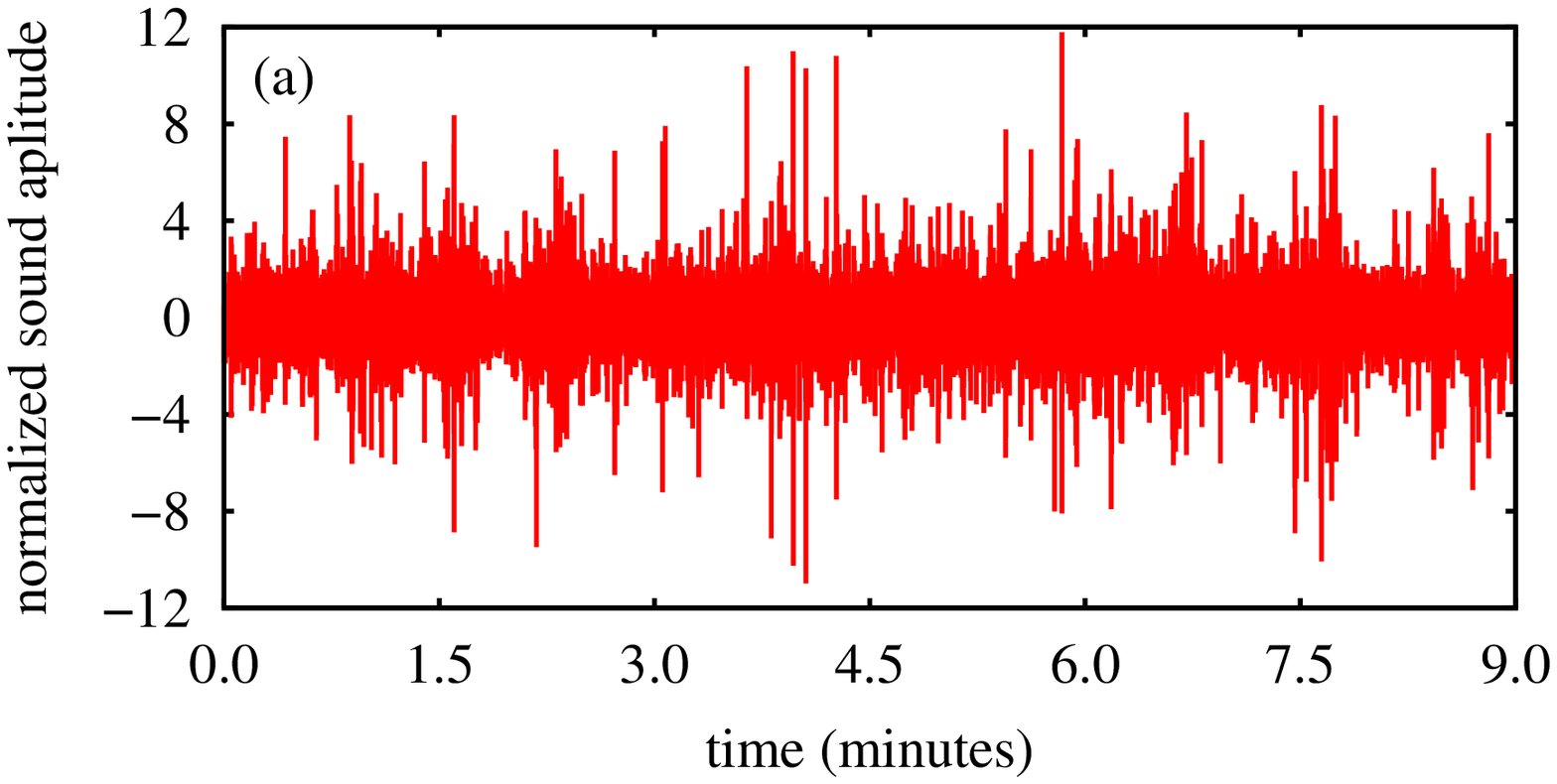}
\includegraphics[scale=0.46]{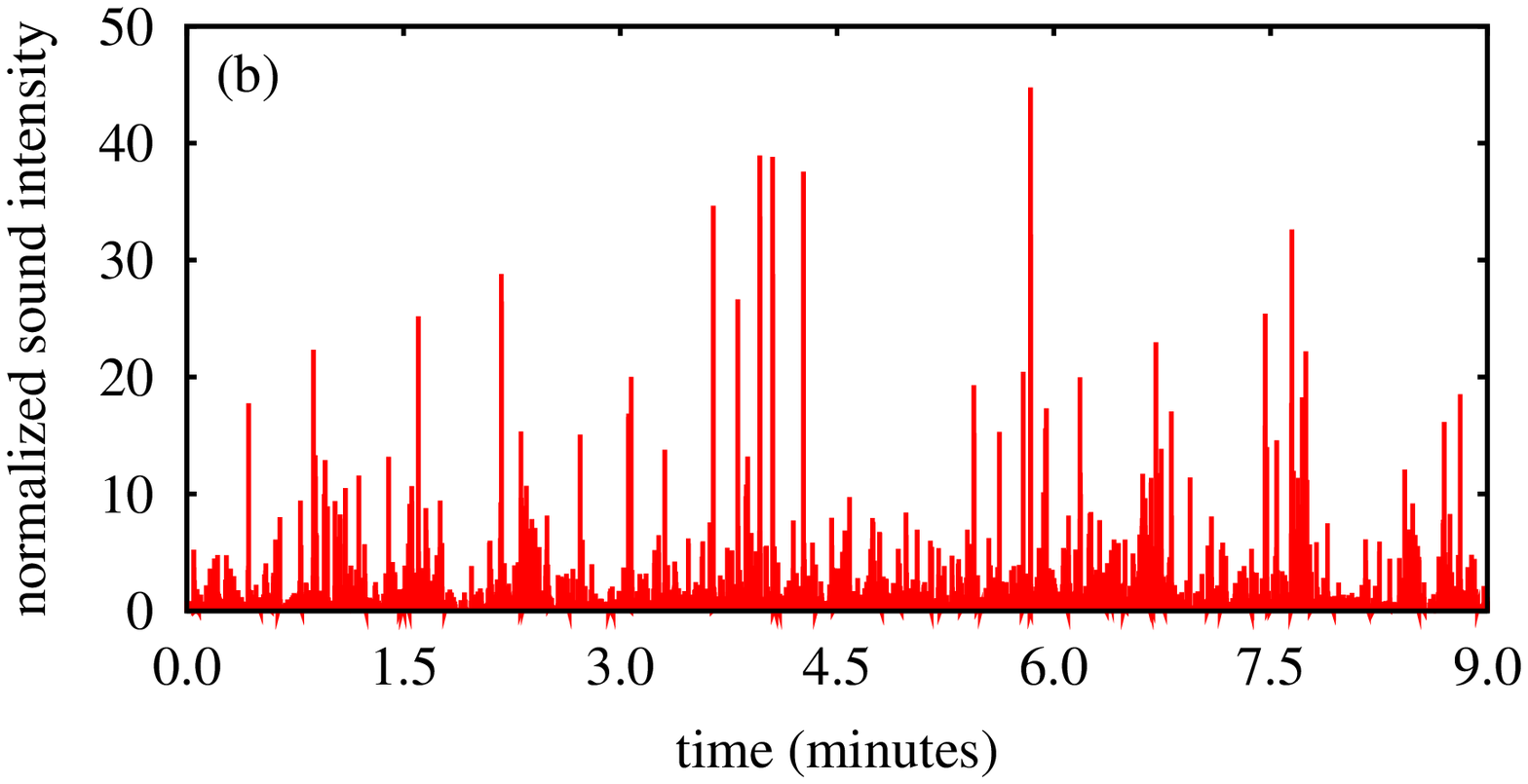}
\caption{(Color online) A representative record sound signal: (a) the normalized sound amplitude and 
(b) the normalized sound intensity.}\label{fig:exp}
\end{figure}
\section{Data presentation}

The observational data was obtained by recording the soundscape of people agglomeration 
in the recreation time at our university. The meeting point is an open place where the students
spend time until the next class. All the measurements were made by using a condenser microphone 
(Shure Microflex MX202W/N) positioned in the central part of the agglomeration. We employed a
sampling rate of 44.1 kHz in order to cover the full audible human range (approximately between 20
Hz and 20 kHz). Additionally, the measurements were made during different periods in nine days
{ totaling 16 records. The number of people during the recordings ranged} approximately between 100 and 200, and these variations 
does not significantly change the statistical results. Typical recording times are about 10 minutes
and along the recording the number of persons was approximately constant. We also analyzed 10 recordings 
from a web sound database\footnote{freesound.org} finding similar results when compared with our measurements. 

Figure \ref{fig:exp}(a) shows a representative record signal where we employed the normalized sound amplitude $A_t$,
i.e., the sound amplitude subtracted by its mean value and divided by its standard deviation. Figure \ref{fig:exp}(b)
presents the sound intensity, $A_t^2$, divided by its standard deviation. From these two figures, we can observe
the existence of some bursts where the sound amplitude  and the sound intensity exceed values much larger than 
their standard deviations. Qualitatively, the origin of these extreme events may be, for instance, related to the fact that the people
want to be heard, and if the neighbors are talking out loud they also have to increase the sound intensity. 
 
\section{Statistical analysis}
\begin{figure}[!ht]
\centering
\includegraphics[scale=0.4]{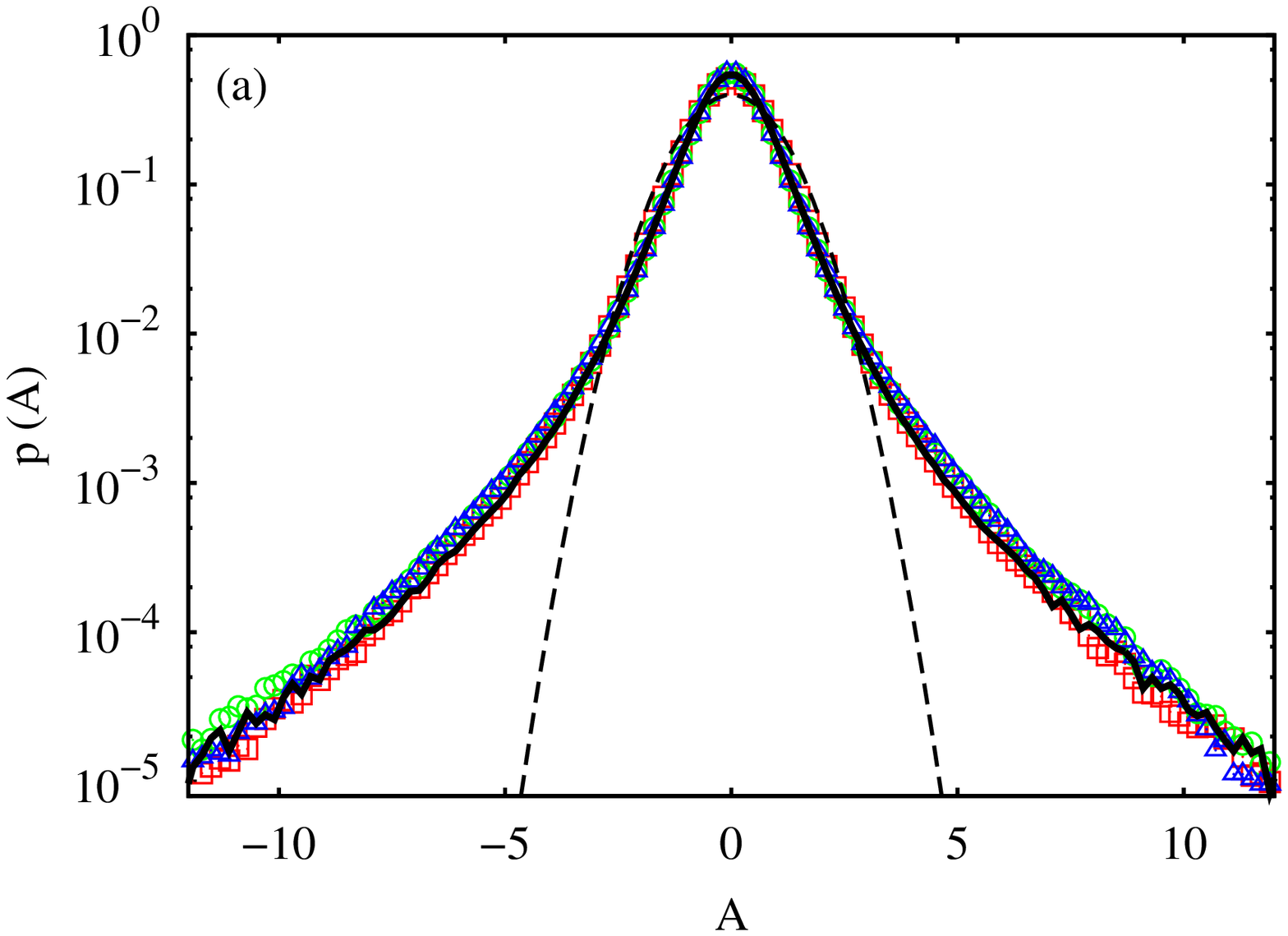}
\includegraphics[scale=0.4]{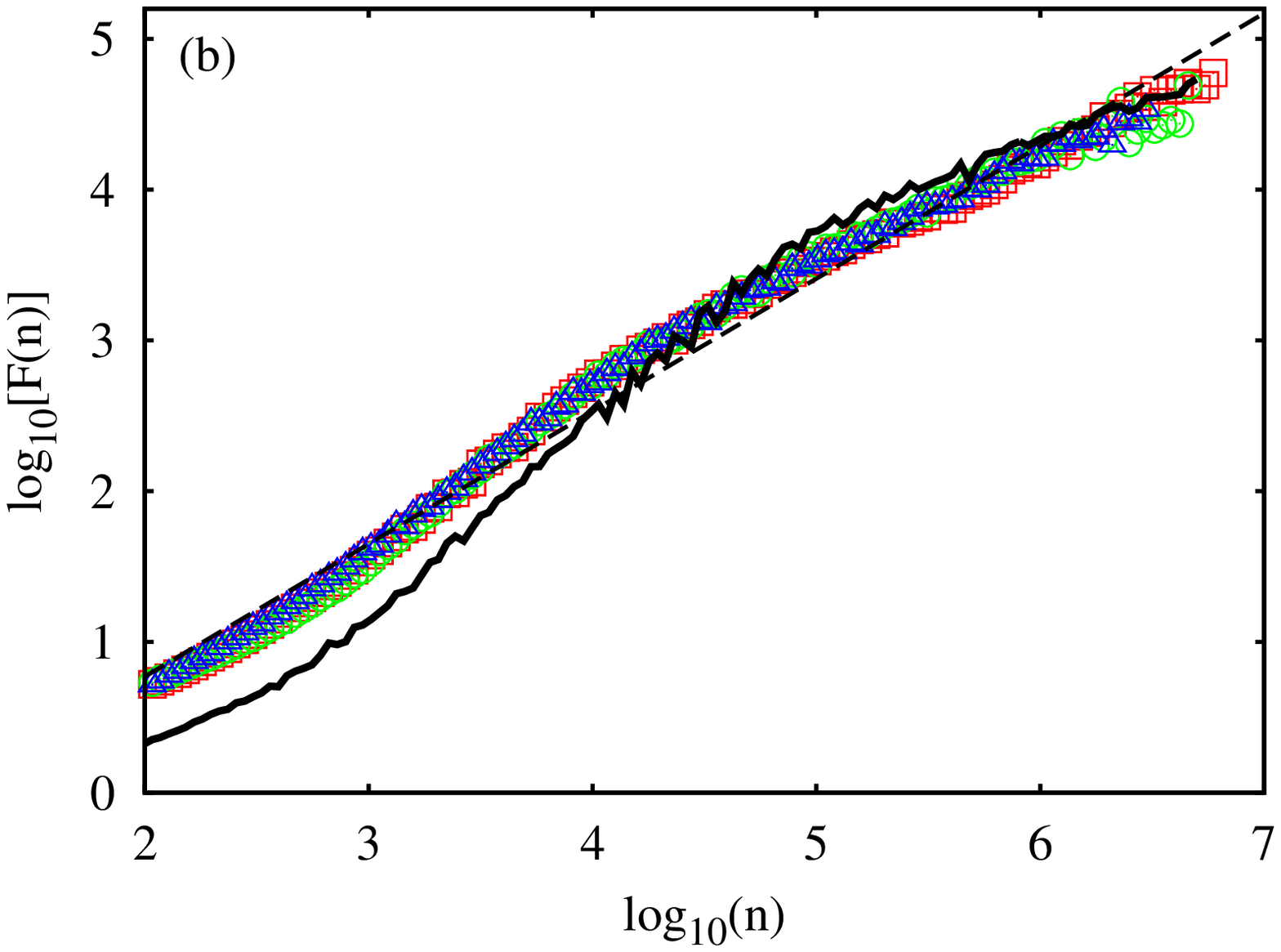}
\includegraphics[scale=0.4]{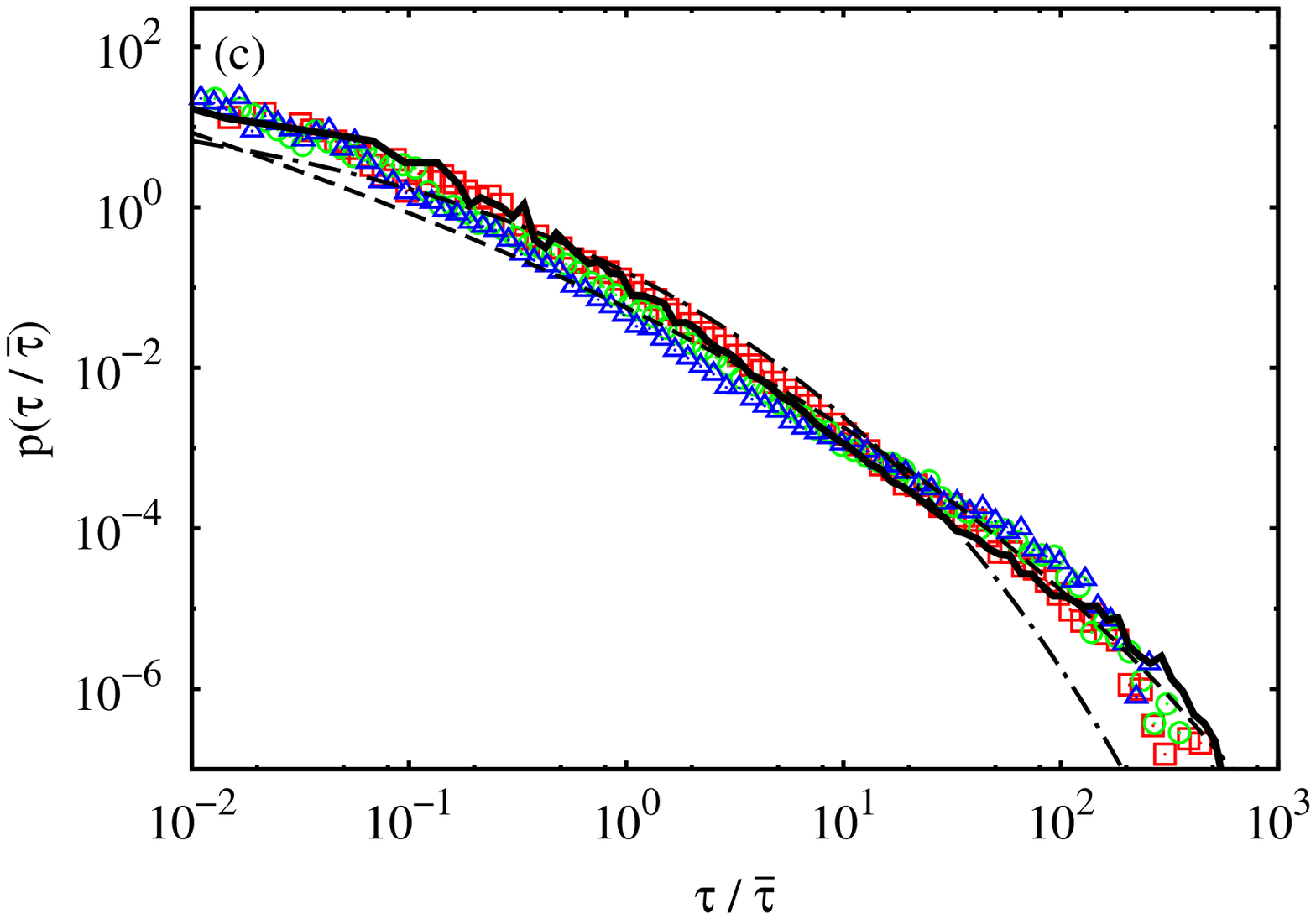}
\includegraphics[scale=0.4]{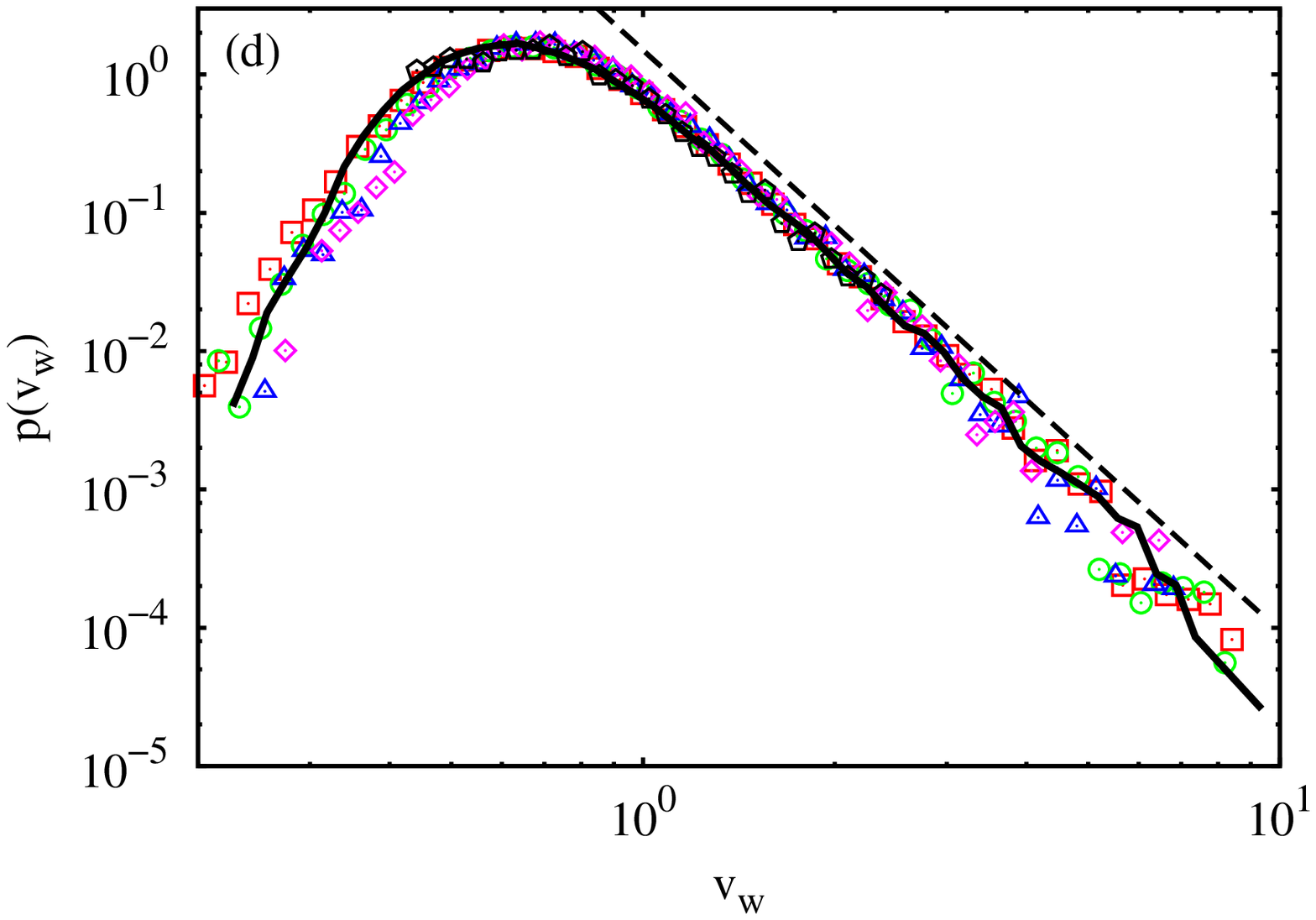}
\caption{(Color online) (a) Probability distribution of the normalized sound amplitude $A$ for three realizations of the experiment
(squares, circles and triangles) confronted with the standard Gaussian pdf (dashed line) and with the GARCH model (continuous line).
(b) DFA analysis when considering the same three previous realizations for the normalized sound intensity: $\log_{10}[F(n)]$ versus $\log_{10}(n)$ in comparison with a linear fit (dashed line), where we found  
$F(n)\propto n^h$ with $h\approx0.88$, and with the GARCH model (continuous line). Here $n$ is in units of 1/44.1$k$ seconds. (c) Return interval distribution take into account one realization of the experiment for three threshold values: $q=1$ (squares),
$q=2$ (circles) and $q=5$ (triangles) compared with the stretched exponential (dashed-doted line) and the Weilbull distribution (dashed line) of equation \ref{eq:stretched}
with $\gamma=2(1-h)=0.24$, and also with the GARCH model (continuous line). (d) Volatility distribution for one realization  of the experiment and considering five window sizes: $w=1$ (squares), 
$w=2$ (circles), { $w=5$ (triangles), $w=10$ (diamonds) and $w=100$ (pentagons)} hundredths of a second. The dashed line is a power law with $p(v)\propto v^{-4.1}$ and the continuous line is the GARCH prediction. }\label{fig:data} 

\end{figure}

One of the most direct ways to characterize the sound amplitude is by evaluating its probability density function  (pdf).
We show this analy\-sis in Figure \ref{fig:data}(a) { for three typical recordings} where we also plot one Gaussian distribution with zero mean and unitary variance (dashed line). 
{ A quite similar behavior has been found for all the other realizations of the experiment and also for the web recordings (at least in the central part of the distribution).}
The empirical distribution clearly differs from the Gaussian one, especially for larger values of the sound amplitude ($|A|$ greater than four standard deviations).  Naturally, this heavy tail behavior reflects the presence of extreme events that we qualitatively see in Figure \ref{fig:exp}.

A possible manner to investigate the dynamics of these extreme events is by evaluating the time interval between them. These
time intervals can be obtained by considering a threshold value $q$ and storing all the initial time $t_i$ for which the normalized 
sound intensity is above this edge. The difference between two consecutive times $\tau_i = t_{i+1}-t_{i}$ is the so called return interval.
For Gaussian uncorrelated (or weak correlated) random variables the distribution of $\tau_i$ is well known to follows an exponential distribution
$p (\tau) \sim  e^{-{\tau}/{\bar{\tau}_q}}$, { where $\bar{\tau}_q$ is the average value of $\tau_i$ when considering the threshold value $q$}. Additionally, empirical results have shown that, in the presence of power law correlations in the data, the distribution is well adjusted by a stretched exponential~\cite{Bunde,Wang,Yamasaki} or by a Weilbull distribution~\cite{Blender}, i.e.,
\begin{equation}\label{eq:stretched}
p(\tau) \sim e^{-A \left({\tau}/{\bar{\tau}_q}\right) ^\gamma} \qquad\mbox{or}\qquad p (\tau) \sim 
\left({\tau}/{\bar{\tau}_q}\right) ^{\gamma-1}e^{-B \left({\tau}/{\bar{\tau}_q}\right) ^\gamma}\,,
\end{equation}
where $A$ and $B$ are constants and $\gamma$ is the exponent of the power law autocorrelation function.  
These distributions also emerge in the analytical framework of Santhanam and Kantz~\cite{Santhanam} when considering 
a long-range correlated noise with Gaussian pdf. Notice that all these distributions are dependent on $q$, but if we employ
the scaled variable  $\tau_i/\bar{\tau}$ this dependence is eliminated. 

Before we investigate the return intervals, let us address the correlation question by using the detrended fluctuation analysis (DFA)~\cite{Peng2}. 
This technique basically considers the root mean square fluctuation function $F(n)$ (see for instance ~\cite{Kantelhardt}) for the integrated 
and detrended time series for different values of the time scale $n$. When the data present scale-invariance properties, $F(n)$ follows a power law 
$F(n)\sim n^h$, where $h$ indicates the degree of correlation in the time series: if $h=0.5$ the series is uncorrelated
and if $h>0.5$  the series is long-range correlated. Figure \ref{fig:data}(b) shows the fluctuation function versus $n$ 
{ for the same three recordings of Figure \ref{fig:data}(a) where we found $h \approx 0.88$ indicating that long-range 
correlations are present in the data. Note that the three curves are practically identical. This fact is evidenced by evaluating
the mean value of $h$ ($\bar{h}$) and its standard deviation ($\sigma_h$) over the 16 recordings finding $\bar{h}=0.88$
and $\sigma_h=0.001$. When considering the web recordings this values remain close: $\bar{h}=0.89$ and $\sigma_h=0.01$.}

Now, advancing with the return interval distribution, it is interesting to emphasize that the exponents $h$ and $\gamma$ are related via $\gamma=2(1-h)$. 
Moreover, since the distribution of $\tau_i/\bar{\tau}$ should be normalized and also have unitary mean, the only fit parameter is $\gamma$ 
that can be obtained from $h$ leading to $\gamma\approx 0.24$. Figure \ref{fig:data}(c) shows this distribution for three values of $q$ where we can observe 
a reasonable data collapse but a not so good agreement with the distributions of equation \ref{eq:stretched}. Similar situation have been recently observed 
when considering non-Gaussian distributions related to water boiling~\cite{Ribeiro2}.

We can also investigate the bursts observed in Figure \ref{fig:exp}(a) by evaluating the volatility of the normalized sound amplitude. 
{ This time series refers to the local standard deviation of $A(t)$ estimated over a time window $w=n \Delta t$, i.e.,}
\begin{equation}
v_w^2(t)=\frac{1}{n-1}\sum_{t'=t}^{t+n-1} (A(t')-\langle A(t)\rangle_w)^2\,,
\end{equation}
{ where $\langle A(t)\rangle_w=\frac{1}{n}\sum_{t'=t}^{t+n-1} A(t')$, $n$ is a integer and $\Delta t$ is the sampling time interval.
Figure \ref{fig:data}(d) shows the volatility distribution of our empirical data for time windows ranging from $1/100$ to $1$ second.
Notice that we found a good collapse of data and that this distribution has an asymptotic power law decay characterized by a exponent $\eta=4.1$.
The mean value and the standard deviation of $\eta$ calculated over the 16 realizations are respectively $\bar{\eta}=4.29$ and $\sigma_\eta=0.35$ 
($\bar{\eta}=4.90$ and $\sigma_\eta=1.10$ for the web recordings).}

\section{Modeling}
Our starting point to model the data behavior is the non-stationary aspect of the volatility. Figure \ref{fig:data}(d) supports 
the conclusion that the volatility of the sound amplitude is a time-dependent stochastic process and Figure \ref{fig:data}(b) 
indicates that long-term memory are present in sound intensity series. This feature
is very common in financial data where the volatility (or risk) is one of most essential ingredients in the price
dynamics. In this scenario, much work has been done~\cite{Mantegna} and consequently a large amount of models are available.
From a qualitative point of view, the interactions (competitions) among people present financial markets seems to be similar to the ones existent 
in our social system. This picture motives us to employ a typical financial model to our data.

One of these models is the generalized autoregressive conditional heteroskedastic processes or simply the GARCH process.
This model was proposed~\cite{Bollerslev} (at least in part) to take into account the long memory typically found in financial data.
It is defined in its most general form, GARCH$(p,q)$,  by
\begin{eqnarray}
x_t&=&\sigma_t \,\xi_t\,,\nonumber\\
\sigma_t^2&=&\alpha_0+\alpha_1 x_{t-1}^2+\dots+\alpha_p x_{t-p}^2+\beta_1 \sigma_{t-1}^2+\dots+\beta_q \sigma_{t-q}^2\,,
\end{eqnarray}
where $\alpha_i$ and $\beta_i$ are positive control parameters and $\xi_t$ is a uncorrelated random variable
with zero mean and unitary variance. Thus, the GARCH process is uncorrelated in $x_t$ but correlated in the
variance. Also note that for $\alpha_i=0$ the GARCH recovery the so called ARCH process~\cite{Engle}.

Here, for simplicity and also for satisfactoriness we will focus on the GARCH$(1,1)$ process
\begin{eqnarray}
x_t&=&\sigma_t \,\xi_t\,,\nonumber\\
\sigma_t^2 &=&  \alpha_0 + \alpha_1 x_{t-1}^2+\beta_1 \sigma_{t-1}^2\,,
\end{eqnarray}
for which we choose the distribution of $\xi_t$ to follow the standard Gaussian. After this simplification the model
have three parameters: $\alpha_0$, $\alpha_1$ and $\beta_1$. However, since the sound amplitude is scaled to a unitary variance, we can
eliminate one of these parameters by using the expected variance of the GARCH$(1,1)$ process $x_t$:
\begin{equation}
\sigma_x^2= \frac{\alpha_0}{1-\alpha_1-\beta_1}\,.
\end{equation}
In this manner, we have now two parameters that we incrementally update to minimize, via the method of least squares, 
the difference between the simulated values of sound amplitude and the observational ones. The best values for the 
parameters are $\alpha_1=0.011$ and $\beta_1=0.9889$ leading to $\alpha_0=0.001$ since $\sigma_x=1$.
The comparison with the empirical data is shown in Figure \ref{fig:data}, where the GARCH$(1,1)$ predictions are 
indicated by the continuous lines. We can see that the agreement between the data and the GARCH$(1,1)$ is very good. 

Concerning Figure\ref{fig:data}(b), where we compare the DFA analysis, we have to remark 
that the autocorrelation function of the variable $x_t^2$ is not really long-range correlated.
In fact, it has an exponential decay~\cite{Bollerslev}, i.e., $\langle x_t^2\, x_{t+\tau}^2\rangle \sim \exp(-t/\tau_c)$, 
where $\tau_c=|\ln(\alpha_1+\beta_1)|^{-1}$. However, the GARCH$(1,1)$ process can mimic the long-range decay
for large values of the characteristic  time $\tau_c$. This feature can be achieved by choosing the sum $\alpha_1+\beta_1$
closer the unity. In our case, $\alpha_1+\beta_1=0.9999$ leading to characteristic  time $\tau_c\sim10^4$ seconds,
which is very large mimicking at least in part the long-range correlations. { Notice that the empirical data also present
deviations from the straight line suggesting that correlations present in the data may have a kind of exponential cutoff.} 

\section{Summary}
In this work we investigated some statistical aspects of the collective sound emitted by people when they are agglomerated
in a meeting place. Empirical evidences showed that (i) the normalized sound amplitude is not Gaussian distributed,
(ii) the sound intensity presents long-range correlations, (iii) the return interval distribution of the sound intensity
is not exponential, and (iv) the volatility of the sound amplitude is non-stationary having a power law tail in its distribution.
Motivated by the time dependence of the volatility, we compared the observational quantities with the predictions
of the GARCH$(1,1)$ model, finding a good agreement with all of them. 

{ Before concluding, we would like to point out some possible mechanisms responsible by the presence of
heavy tail distributions and long-term correlations in the data. The first one is related to the fact that humans 
already have an intrinsic complex behavior which may manifest in our measurements. Second, these individuals form small 
interacting groups adding more complexity to the system. On a third level, there is also an emergence of interactions
between groups. Naturally, more detailed measurements and models should be considered, in comparison with those
one presented here, to obtain a broad understanding of this system.
}

%\section{acknowledgements}
\ack
We thank CNPq/CAPES (Brazilian Agencies) for financial support and CENAPAD-SP for computational support.
L.R.E. and R.T.S. also thank CNPq/INCT-FCX for the financial support.

\end{document}